\documentclass[prd,singlecolumn,superscriptaddress,showpacks]{revtex4}
\usepackage[utf8]{inputenc}
\usepackage[T1]{fontenc}
\usepackage{empheq}
\usepackage{amsmath}
\usepackage{amssymb}
\usepackage{stmaryrd}
\usepackage{dsfont}
\usepackage{graphicx}
\usepackage{hyperref}
\usepackage{url}
\usepackage{stackrel}
\usepackage{color}
\usepackage{comment}
\usepackage{mhchem}
\usepackage{calc}
\usepackage{cleveref}
\usepackage[shortcuts]{extdash}

\def \ee{\end{equation}}
\def \be{\begin{equation}}
\def \eea{\end{eqnarray}}
\def \bea{\begin{eqnarray}}

\newcommand{\up}{\hspace{-0.5ex}\uparrow}
\newcommand{\down}{\hspace{-0.5ex}\downarrow}
\newcommand{\diagram}[2][0.5]{
  \raisebox{0.5ex-#1\height}{\includegraphics{#2}}
}
\newcommand{\Eqref}[1]{Eq.~\eqref{#1}}

\begin{document}

\title{
Exciting hint toward the solution of the neutron lifetime puzzle
}
\author{Benjamin Koch}
\email{benjamin.koch@tuwien.ac.at}
\affiliation{Institut f\"ur Theoretische Physik,
 Technische Universit\"at Wien,
 Wiedner Hauptstrasse 8--10,
 A-1040 Vienna, Austria}
\affiliation{Instituto de F\'isica, Pontificia Universidad Cat\'olica de Chile, 
Casilla 306, Santiago, Chile}
\author{Felix Hummel}
\affiliation{Institut f\"ur Theoretische Physik,
 Technische Universit\"at Wien,
 Wiedner Hauptstrasse 8--10,
 A-1040 Vienna, Austria}
\date{\today}

\begin{abstract}
We revisit the neutron lifetime puzzle, a discrepancy between beam and bottle measurements of the weak neutron decay.
Since both types of measurements are realized at different times after the nuclear production of free neutrons, we argue that the existence of excited states could be responsible for the different lifetimes.
We elaborate on the required properties of such states and under what
circumstances it is possible that it has not been experimentally identified yet.
\end{abstract}

\maketitle

\tableofcontents

\section{Introduction}

\subsection{The neutron in the quark model}

The neutron is one of the main constituents of nuclear matter.
It is
a composite state, whose properties are ruled by the strong and the electroweak interactions between the lightest quarks of the Standard Model of particle physics.
It is of pivotal importance in many phenomena 
ranging from Big Bang Nucleosynthesis to
experimental particle physics~\cite{ParticleDataGroup:2022pth}.

Even though, a detailed understanding of the low energy properties of this particle in terms of fundamental degrees of freedom is an open field of research,
it is possible to understand several properties in terms of much simpler models.
In our discussion we will make use of the language and notation of the quark model.
In this model, protons and neutrons are composite particles made up of quarks. Protons consist of a particular combination of two ``up'' ($u$) quarks and one ``down'' ($d$) quark, while neutrons consist of combinations of one up quark and two down quarks. While quarks carry a fractional electric charge, the combination of quarks in protons and neutrons results in particles with integer electric charges.

Isospin describes the similarity between protons and neutrons. It
was introduced by Werner Heisenberg and later developed further by
Eugene Wigner~\cite{Heisenberg:1932dw,Wigner:1936dx}.
Algebraically, the isospin operator $\vec I$, can be represented analogously to the spin operator of spin one half particles $\vec S$ in terms of the Pauli matrices $\vec \sigma$, which are a representation of the group $SU(2)$.
In the quark model it is imposed that both $\vec I$ and $\vec S$
are (approximate) symmetries. 
Thus actions of the group memebers are to be understood as symmetry transformations.
Based on this,
one imposes that the neutron wave function is an eigenfunction of $(\vec I^2, \,I_z)$ with the eigenvalues $(3/4 , \, -1/2)$ while the proton has the eigenvalues $(3/4, +1/2)$.
Further imposing that these particles carry spin $\pm 1/2$ like their three constituents singles out a state for the neutron (e.g. with spin up $\uparrow$~\cite{GreinerMueller})
\be
|n \up\rangle=\frac{1}{3\sqrt{2}}
\left( |udd\rangle |dud\rangle |ddu\rangle\right)\left(
\begin{array}{rrr}
     2&-1  &-1\\
-1     & 2&-1\\
-1&-1&2
\end{array}\right)
\left(
\begin{array}{c}
     |\down \uparrow \uparrow\rangle \\
     | \up \downarrow\uparrow\rangle \\
     |\up \uparrow \downarrow \rangle \\
\end{array}
\right).
\label{eqn:ground_state_flavor_spin}
\ee
The corresponding unique state for the proton is obtained analogously by replacing $u \leftrightarrow d$.
If the isospin symmetry were an exact symmetry protons and neutrons would have the same energy and thus the same mass. 
The observed difference between the neutron mass $m_n$ and the proton mass $m_p$ and the aforementioned neutron decay show that $I_z$ must be broken. 
In the above description, this breaking is modeled by associating a larger mass to up quarks than to down quarks $m_u<m_d$.
This generates a mass splitting between $|n \up\rangle$ and $|p \up\rangle$.
Also the conservation of spin is only an approximate concept, due to the presence of gluons, virtual quark antiquark pairs, and the respective angular momenta. Interestingly, the presence of these virtual particles can be effectively absorbed in the concept of constituent quarks as ``dressed'' color states~\cite{Lavelle:1995ty} which combine such that they form color neutral hadrons.
This, and other, more sophisticated models of the neutron, in combination with  experimental efforts allowed to learn more and more details about 
the properties such as mass, composition, and lifetime~\cite{ParticleDataGroup:2022pth}. Thus, it is fair to say that quark models have proven to be useful for our understanding of the internal structure of hadrons~\cite{Eichmann:2016yit}.

Since we will exemplify our ideas with a model which has also  three quarks, a word of caution is in place here. 
There are several phenomenological aspects which can hardly be captured even by our most sophisticated theoretical models, not to speak
by simple three\-/quark models: 
\begin{itemize}
    \item[(i)] Mass:\\ 
    The simple quark model fails when one attempts to predict the precise mass of hadrons~\cite{GreinerMueller}. 
    \item[(ii)] Spin:\\
    The structure of the spin distribution inside of a neutron wave\-/function is largely unknown (for a review see \cite{Deur:2018roz}). However, there is certain agreement that the valence quarks only carry a fraction of the neutron spin, which was at times labeled as ``spin crisis'' \cite{Veneziano:1989ei,Anselmino:1988hn,Nayak:2018nrv}.
    \item[(iii)] Radius:\\
    The radius of the proton has been deduced from complementary measurements and the resulting $4\%$ disagreement became widely known as the ``proton radius puzzle''~(for a review see \cite{Carlson:2015jba}).
    \item[(iv)] Neighbors:\\
    The inner structure of protons and neutrons seems to be strongly sensitive to the neighboring hadrons within a nucleus. This puzzling behaviour is known as ``European Muon Collaboration effect'' \cite{Geesaman:1995yd,Koch:2019qqw}.
    \item[(v)] Lifetime:\\
    Last but not least, there is a tension between different measurements of the neutron's lifetime~(see e.g. \cite{Paul:2009md,Serebrov:2011re}).
\end{itemize}

	Thus, any three\-/quark model should be understood as useful tool for qualitative understanding rather than for precise quantitative modelling.

The last item on the above list will be the subject of interest of this paper. It will therefore be summarized in the following subsection.

\subsection{The lifetime puzzle}

In this study, we will focus on the discrepancy in the lifetime of the neutron, known as the Neutron Lifetime Puzzle (N$\lambda$P):

Beam neutrons have an about 10 sec longer lifetime than Ultra Cold Neutrons  (UCN) in bottle traps
~\cite{Paul:2009md,Wietfeldt:2011suo,Ezhov:2014tna,Pattie:2017vsj,Serebrov:2018yxq,UCNt:2021pcg,Hirota:2020mrd}.
The corresponding normalized difference for UCNs 
in gravitational bottle traps and average beam results is
%
%
\be\label{eq_puzzle}
\Delta \tau=\tau_\mathrm{beam}-\tau_\mathrm{bottle}=(8.6\pm 2.2) \,\mathrm{s}.
\ee
One expects this quantity to be compatible with zero and the fact
that it is significantly different from zero is known as the N$\lambda$P.
This puzzle has persisted over the years and
a possible explanation due to exotic decay channels was explored~\cite{Klopf:2019afh,Dubbers:2018kgh,aSPECT:2008vas,Fornal:2023wji}, but the status is not conclusive yet.
An alternative theoretical 
conjecture is based on neutron oscillations~\cite{Berezhiani:2018eds,Berezhiani:2018xsx,Tan:2023mpj,Altarev:2009tg}. Also this possibility has, been constrained~\cite{Broussard:2021eyr}. A recent proposal suggests that the discrepancy could be understood through the inverse quantum Zeno effect \cite{Giacosa:2019nbz}.
Further conjectures include Kaluza--Klein states \cite{Dvali:2023zww}, modifications of the CKM matrix induced at the TeV scale~\cite{Belfatto:2019swo},
and dark photons \cite{Barducci:2018rlx}.
Also concerns about the systematic error in beam experiments have been raised~\cite{Serebrov:2020rvv}, 
but a contradicting response was given in~\cite{Wietfeldt:2023nlh}.
Thus, the status of how to interpret, or understand the N$\lambda$P \Eqref{eq_puzzle} is still inconclusive.
What makes this problem even more interesting is the fact that it is not only a pressing problem for theoretical model building, it is a disagreement between two complementary types of experiments. Clearly, it would be helpful to have more results with beam experiments, but meanwhile 
we opt to trust the results of our experimental colleagues. Thus, a solution of this discrepancy is urgently needed.

\subsection{Structure of the paper}
In the following Section \ref{sec_Hypo} we present our working hypothesis and formulate necessary conditions for a solution of the N$\lambda$P by means of excited states.
In Section \ref{sec_ToyModel} we provide a toy model that is able to fulfill most of these necessary conditions. In Section \ref{sec_Discussion} we explore the parameter space of the toy model and discuss observational issues and known analogies in physics. Conclusions are given in Section \ref{sec_Conclusion}.

\section{Excited states hypothesis}
\label{sec_Hypo}

Beam neutrons and UCNs are very similar with respect to most important characteristics.
There are, however, some interesting differences. 
First of all there is the difference in velocity.
While neutron beams operate typically with velocity at the order of ${\mathcal O}(10^3\,\mathrm{m/s})$,
neutrons in bottles have velocities of ${\mathcal O}(10^0\,\mathrm{m/s})$. 
Another distinction is time. Beam neutrons
are measured very shortly after their production in the reactor.
The time scale for this is of the order of $t_\mathrm{beam}\sim 2 - 25 $\,ms.
Bottle neutrons, instead, have to undergo a process of slowing,
orienting, and cleaning, which means that their weak decay is measured
a long time after their production $t_\mathrm{bottle}> 300\,\mathrm{s}$~\cite{UCNt:2021pcg}.

\begin{figure}[hbt]
   \centering
\includegraphics[width=10cm]{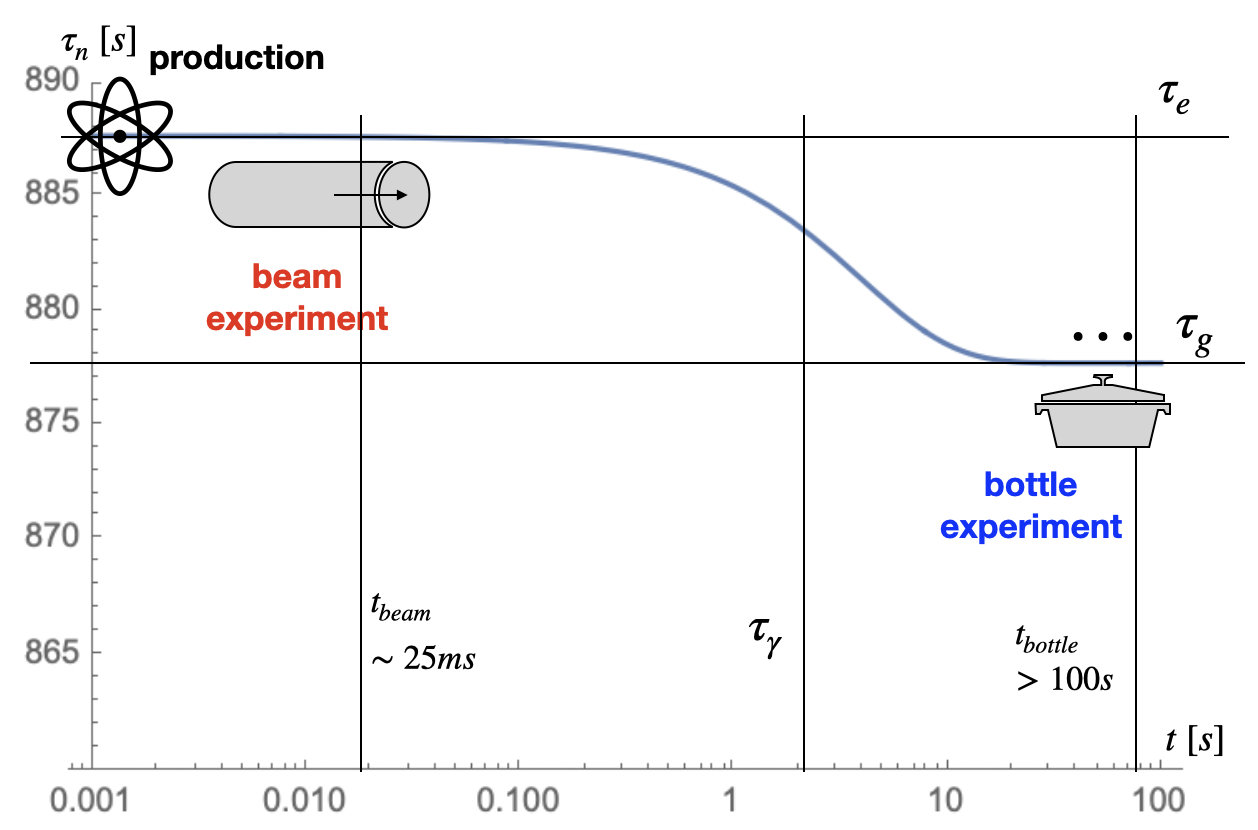}
\caption{\label{fig_Lt}
Composition\-/averaged lifetime \Eqref{eq_taueff} as a function of the time for an exemplary value of $\tau_\gamma =4\,$s.
 }
\end{figure}

This difference is the motivation for our proposal.
Let us assume that the wave functions of free neutrons can have excited states
which are inaccessible for neutrons within a strongly bound hadronic ensemble.
Let us further assume that there is a ground state $\psi_g$ and
excited states $\psi_e$
with different lifetimes $\tau_g\neq\tau_e$ under beta decay. 
These states of the free neutron shall be connected by an electromagnetic
channel $\psi_e \to \psi_g + \gamma$ with a cumulative decay time $\tau_\gamma$. Note that there are could be multiple excited states. However, to keep the following discussion simple, we treat 
$\psi_e$ as a representative of these excited states. A generalization to multiple states is then straight forward.
Disregarding possible byproducts, we propose the following decay cascade:
\be\label{eqn_cascade}
\diagram{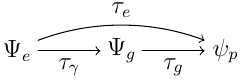}
\ee
From \Eqref{eqn_cascade} the decay rates of the corresponding particle numbers in these states are
\bea\label{eq_dotn}
\dot n_e&=& -n_e\tau_\gamma^{-1} - n_e\tau_e^{-1},\\ \nonumber
\dot n_g&=& +n_e\tau_\gamma^{-1} - n_g\tau_g^{-1}.
\eea
If the proposed excited states and the ground state are separated only
by a comparatively small energy the decay $\psi_e\to\psi_g$ could happen
unnoticed by current experimental setups and
both states of the free neutron would be counted together
\be
n_n=n_g + n_e.
\ee
Note that the two proposed states of the free neutron may be experimentally
distinguishable for certain setups.
We elaborate on that in Section~\ref{sec_Discussion}.
The lifetime of the mixture would be extracted from
\be\label{eq_taueff}
\tau_n=-n_n\dot n_n^{-1}.
\ee
If the characteristic times for the transitions are ordered as follows
\be\label{eq_ltgamma}
t_\mathrm{beam}(\approx 2\times 10^{-2}\,\mathrm{s})\ll \tau_\gamma\ll t_\mathrm{bottle}<\tau_g<\tau_e
\ee
it is possible that neutrons in the beam have a longer lifetime
than neutrons after the cooling process, as measured in the bottle experiments.
As an example, Fig.~\ref{fig_Lt} plots the total composition\-/averaged lifetime $\tau_n$
of neutrons in either state as a function of time
from \Eqref{eq_dotn}, where we arbitrarily chose
$\tau_\gamma=4 \,$s and $n_e(0)=1, \; n_g(0)=0$.
%
The initial abundancies $n_e(0)$ and $n_g(0)$ provide additional degrees of freedom. This occurs because varying initial abundances results in different lifetimes as determined by beam experiments.
For the discussion of the initial abundancies we define the difference of
the beta decay lifetimes of the excited and the ground state as
\be
    \Delta \tau_e=\tau_e-\tau_g,
\ee
which may be larger than the observed difference in lifetime between
the beam and the bottle experiments.
Since we propose $t_\mathrm{bottle}\gg \tau_\gamma$, all neutrons in the bottle are
in the ground state. Thus,
\be
  \tau_n(t_\mathrm{bottle}) = \tau_\mathrm{bottle} = \tau_g.
\ee
On the other hand, since we also propose that $t_\mathrm{beam}\ll$ all other times,
we can identify the beam abundances as the initial abundances $n_i(0)$.
Thus, $\tau_n(0)= \tau_\mathrm{beam} = \tau_\mathrm{bottle} + \Delta \tau$.
Using \Eqref{eq_taueff} one can solve for the initial relative
abundances that have the observed lifetime in the beam
\be\label{eq_n2ini}
  \frac{n_e(0)}{n_g(0)}= \frac{\Delta \tau}{\Delta \tau_e-\Delta \tau}+
  \mathcal{O}\left(\left(\tau_\mathrm{bottle}^{-1}\Delta \tau\right)^2\right).
\ee
From this relation one realizes that $\Delta \tau_e\ge \Delta \tau$
and that for the extremal case $\Delta \tau_e \rightarrow \Delta \tau$
all neutrons need to be in one of the excited states $n_e(0)\rightarrow 1$.
Let us now briefly summarize the necessary ingredients that we identified to make the proposal work:
\begin{itemize}
\item[a)] 
There is at least one excited state of the free neutron $\psi_e$ above the ground state $\psi_g$.
\item[b)]
The lifetime of the excited states under beta decay is larger than, or equal to,
the neutron lifetime measured in the beam.
\item[c)]
The lifetime  for the electromagnetic transition $\psi_e \rightarrow \psi_g+\gamma$
lies between the time\-/scales of beam and bottle experiments
($t_\mathrm{beam} <\tau_\gamma< t_\mathrm{bottle}$). This  hierarchy of lifetimes
given in \Eqref{eq_ltgamma} is sketched in Fig.~\ref{fig_LtHie}.
\begin{figure}[hbt]
   \centering
\includegraphics[width=10cm]{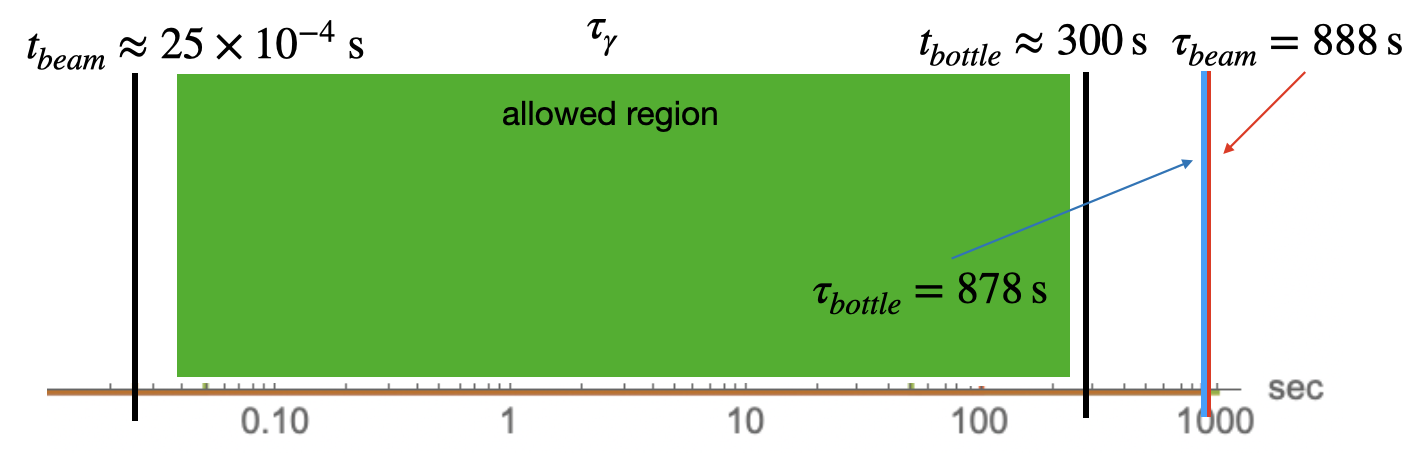}
\caption{\label{fig_LtHie}
Schematic sketch of the different time-scales involved in the N$\lambda$P.
 }
\end{figure}
\item[d)]
The energy difference between $\psi_e$ and $\psi_g$ is smaller than $\sim 0.1 - 10\,$MeV, such that the $\psi_e$ can be excited with the initial kinetic energy $\Delta E$ available from nuclear fission. Further constraints
on the energy difference will be developed throughout this paper.
\item[e)]
When a free neutron is created from a nuclear reaction, the initial abundancy of the excited states follows \Eqref{eq_n2ini}
to yield the measured effective beta decay lifetime in the beam.
\item[f)]
The states $\psi_e$ are not populated if the neutron is strongly bound within a nucleus.
Accurate measurements involving strongly bound neutrons~\cite{ParticleDataGroup:2022pth} in nuclei disfavor the population
of a excited states within the nucleus and in reactions with this bound state.
\end{itemize}
The items (a--f) are necessary conditions for an alternative solution of the N$\lambda$P.
How could this be realized?
Differences in lifetimes under $\beta$ decay of 
states with identical couplings at the quark level are conceivable in two different ways:
Either an enhanced phase space or selection rules of quantum numbers.
On the one hand the decay of states with larger energy have a larger phase space,
leading to a reduced lifetime of excited states.
This is the opposite effect of what is needed
to address the N$\lambda$P.
However, for sufficiently small energy differences between
the states of the free neutron, the effect
of the enhanced phase space is negligible.
On the other hand, excited states may have
quantum numbers that partially disfavor the $\beta$ decay
compared to the ground state. Such selection rules can strongly
lower the decay rates.

To further explore the
latter option, we will recur
to a simple toy model, where a neutron is built of
three quarks $udd$, which are assumed to carry its total spin,
while neglecting the contributions of internal angular momentuma and
virtual particles.

\section{A model with three quarks} 
\label{sec_ToyModel}

The purpose of this section is to give an idea of how the above hypothesis can be implemented in terms
of a simple but concrete model.
Let us assume that a neutron consists of three non\-/relativistic constituent quarks $udd$
The breaking of the isospin symmetry is
only responsible for the explanation of the neutron as an excited state of the proton.
It is not sufficient to implement the hypothesis of an excited neutron state
far below the Roper resonance. For this resonance in the quark model see for
instance Ref.~\onlinecite{Eichmann:2016yit}.

Here we want to extend the constituent quark model of
\Eqref{eqn:ground_state_flavor_spin} by spatial degrees of freedom
to model possible excited states.
The total wave function must be anti\-/symmetric under exchange of any of the quarks.
Color confinement requires
anti\-/symmetry in the color quantum numbers and allows the color part of
the wave function to be factored out for the constituent quark model~\cite{Han:1965pf}.
The remaining flavor, spin, and spatial part must therefore be symmetric under
the exchange of any two quarks.

Although isospin is an approximate symmetry, we assume that neutron and proton
states before and after the decay are total isospin eigen states.
We also suppose that neutron and proton states are total spin eigen states
with a total spin of 1/2.
Excited states of the neutron with a total spin $3/2$ are
known as $\Delta$\-/baryons, however, they are far heavier than the neutron
itself~\cite{ParticleDataGroup:2022pth}.

\subsection{Spatial wave function}
We do not attempt to fully model the spatial part of the wave function.
Quantitative models exist for heavy mesons, such as
charmonium or bottomonium~\cite{ParticleDataGroup:2020ssz}.
Here, we only want to assess the necessary conditions for the hypothesized
excited states of the neutron.
We expand the wave functions in terms of products of one\-/body
wave functions denoted by
$|\vec i,\vec j,\vec k\rangle=\phi_{\vec i}(\vec r_1)\phi_{\vec j}(\vec r_2)\phi_{\vec k}(\vec r_3)$
where $\phi_{\vec i}$ are the normalized spatial wave functions.
All spatial coordinates are in the restframe of the center of gravity.
The functions $\phi_{\vec i}$ are mutually orthogonal to each other and ordered
in the kinetic energy. The quantum numbers of each quark
$\vec i,\vec j,\vec k$ are each triples $(i_1,i_2,i_3)$, denoting the number of nodes
in each cartesian direction.
The wave function $\phi_{\vec 0}$ has no spatial nodes and the lowest
kinetic energy contribution in an effective confining potential.
Assuming an harmonic-oscillator-like confinement,
the kinetic energy of the wave functions $\phi_{\vec i}$
increases linearly with the number of nodes in the three cartesian coordinates.
Here, we only consider the spatial product state $|\vec 0,\vec 0,\vec 0\rangle$ for the ground
state of the neutron and a general spatial product state $|\vec i,\vec j,\vec k\rangle$
with $\vec i \neq \vec j\neq \vec k\neq \vec i$.
Due to permutational symmetry, all six permutations of $(\vec i,\vec j,\vec k)$
also have to be considered.

If the effective confining potential were fixed to the extent of
the neutron inside the nucleus
the kinetic energy difference between $\phi_{\vec 0}$ and $\phi_{1 \hat z}$ with one
spatial node would be large.
However, we know that a free neutron
can acquire considerable extent outside the nucleus,
lowering the kinetic energy differences with increasing extent.
Assuming for instance a spatial extent of 1\,\AA\ of non-relativistic constituent quarks
with a mass of about 330\,MeV, each, the energy difference between the states
$|\vec 0,\vec 0,\vec 0\rangle$ and $|1\hat z,\vec 0,\vec 0\rangle$ would merely be 40\,meV,
the thermal energy at room temperature.

As detailed in Section \ref{sec:electromagnetic_decay}, we expect
the total number of nodes to be large and depending of the type of the decay
up to the order of $10^2$ to $10^6$
with an equally long decay cascade via electromagnetic multipole transitions.
The total energy difference between $|\vec i,\vec j,\vec k\rangle$ and $|\vec 0,\vec 0,\vec 0\rangle$,
and hence that of the ground and the highest of the excited states of the neutron is given by $\Delta E$.

\subsection{Weak decay}

The weak decay of the neutron is mediated by a $W^-$ boson with spin one.
This boson couples to the down quarks, which have their respective spin.
Given the effective wave functions of the neutron and the proton ($\psi_n, \, \psi_p$),
the main part of this coupling is described in terms of the Lagrangian~\cite{Weinberg:1958ut}
\be\label{eq_LagrWeak}
{\mathcal{L}}=-\frac{G_F}{\sqrt{2}}W_\mu \bar \psi_p \gamma^\mu(g_V + g_A \gamma^5)\psi_n + {\mathcal{O}}\left(\dots \right)_{\mathrm{WM},\mathrm{S}},
\ee
where $G_F$ is the Fermi coupling, $W_\mu$ is a weak current vector, $g_V$ is the vector coupling, $g_A$ axial coupling with the ratio $\lambda = g_A/g_V=-1.275$.
In \Eqref{eq_LagrWeak} we omitted writing explicitly  the subleading higher order corrections,  induced by weak magnetism (WM) between the outgoing states, and the typically vanishing induced scalar terms (S) (for a review see e.g.~\cite{Fornal:2023wji,Byrne:2002as}).

This Lagrangian is, however, insensitive to the internal structure of the neutron, except of the values for the phenomenological parameters $g_V$ and $g_A$.
An accurate model for the weak-decay treating the internal quark structure is
a difficult task of Quantum Chromodynamics (QCD) and beyond the scope of this work.
Here, we want to come up with a qualitative model of the weak decay interaction
on the level of constituent quarks, the level where we model the internal
structure of excited neutrons.
To keep the model as simple as possible, we assume the following:
\begin{itemize}
\item[(i)] All momenta are purely non-relativistic.
Thus, spin conservation and orbital angular momentum conservation separate
and there is no coupling between the two spin channels.
\item[(ii)] The weak decay couples only to a single $d$ quark in the constituent quark model.
\item[(iii)] We model only the down-to-up quark transition, as we are only interested
in contractions between proton and neutron states occurring in
Fermi's golden rule for decay rates.
\end{itemize}
Following these guidelines, we model the following interaction Hamiltonian:
\begin{align}
\nonumber
\hat H_\beta = g_\beta\, |u\rangle\langle d|
\otimes\Big(
  &+\sqrt{\frac23}\big(
    |1,+1\rangle_W|\down\rangle\langle\uparrow\hspace{-.5ex}|
    -|1,-1\rangle_W|\up\rangle\langle\downarrow\hspace{-.5ex}|
  \big) \\
  &-\sqrt{\frac13} \big(
    |1,+0\rangle_W|\up\rangle\langle\uparrow\hspace{-.5ex}|
    -|1,+0\rangle_W|\down\rangle\langle\downarrow\hspace{-.5ex}|
  \big)
\Big)
\otimes \hat 1
\label{eqn:weak_interaction_model}
\end{align}
acting on the flavor, spin, and spatial part of the model wave function.
Here, $|1,s_z\rangle_W$ denotes the emitted $W^-$ boson which can come with
the spin quantum numbers $s_z\in\{-1, 0,+1\}$.
The corresponding factors in the spin--spin part of the interaction are determined from
spin angular momentum conservation of initial and final states.

Taking the spatial degrees of freedom into account,
the (non\-/normalized) ground state of the neutron of \Eqref{eqn:ground_state_flavor_spin}
reads
\begin{equation}
|g,\uparrow\rangle_n = \mathcal S\left\{\big(
    2|ddu\rangle-|dud\rangle-|udd\rangle
  \big) \times \big(
    2|\up\uparrow\downarrow\rangle
    - |\up\downarrow\uparrow\rangle - |\down\uparrow\uparrow\rangle
  \big)
  \times |\vec 0,\vec 0,\vec 0\rangle
  \right\},
\end{equation}
where the symmetrization operator $\mathcal S\{\cdot\}$ sums over all 6 permutations
of the three constituent quarks.
Without loss of generality, we align the neutron with its spin pointing in
the $+z$ direction.
Notice that the symmetrized wave functions of the considered
states do not in general separate into products of flavor, spin, and spatial
wave functions as the symmetrization acts after the tensor product.
We now construct an excited state of the neutron in our model
by replacing the spatial part of the wave function
$|\vec 0,\vec 0,\vec 0\rangle = \phi_{\vec 0}(\vec r_1)\phi_{\vec 0}(\vec r_2)\phi_{\vec 0}(\vec r_3)$ by
the wave function
$|\vec i,\vec j,\vec k\rangle  = \phi_{\vec i}(\vec r_1)\phi_{\vec j}(\vec r_2)\phi_{\vec k}(\vec r_3)$
with $\vec i\neq \vec j\neq \vec k\neq \vec i$.
The set of all total isospin and total spin neutron eigenstates together
with all six permutations of $(\vec i,\vec j,\vec k)$
non-trivially spans a four dimensional space due to permutational symmetry
and we denote any four orthonormal basis vectors of the span by $|b_m,\uparrow\rangle_n$
with $m\in\{1,\ldots, 4\}$.
All calculations concerning the weak interaction Hamiltonian
have been aided by a computer algebra system.

We model the internal structure of the 
proton states after a decay by
replacing all down quarks by up quarks and vice versa.
The spatial structure of the proton is
assumed to be the same as in the neutron before the weak decay.
Following the weak decay, the excited proton is expected to continue its
electromagnetic decay into its ground state.
The spin angular momenta of the final
proton and the $W^-$ boson must add up to the spin angular momentum
of the initial neutron.
Following the above construction for a neutron in the state $|\psi,\uparrow\rangle_n$,
that was initially aligned in the $+z$ direction,
the respective spin-conserving final state
of the proton together with the $W^-$ boson is thus
\begin{equation}
  |\psi\rangle_{p}|\cdot\rangle_W
  = \sqrt{\frac23}\,|\psi,\downarrow\rangle_p|1,+1\rangle_W
  - \sqrt{\frac13}\,|\psi,\uparrow\rangle_p|1,0\rangle_W.
\end{equation}

We now project the modeled (non\-/hermitian) weak interaction Hamiltonian
$\hat H_\beta$ from the left and the right onto the span of the constructed
excited proton and neutron basis states, respectively:
\begin{equation}
  \left(\tilde H_\beta\right)_{lm} = \langle b_l|_p\, \hat H_\beta|b_m,\uparrow\rangle_n.
\end{equation}
Due to spin angular momentum conservation of $\hat H_\beta$
and the neutron being initially in a spin up state,
we can omit
the $W^-$ boson states in the above contractions.
This projection yields a symmetric $4\times4$ eigenvalue problem
with the eigenvalues $\{5, 1, 1, -3\}g_\beta/9$.
The neutron states to the eigenvalues $5g_\beta/9$ and $-3g_\beta/9$ are
for instance
\begin{align}
  \nonumber
|e^+,\uparrow\rangle_n =&\ \mathcal S\bigg\{\big(
    2|ddu\rangle-|dud\rangle-|udd\rangle
  \big)
  \\
  & \times \big(
    2|\down\uparrow\uparrow\rangle
    - |\up\uparrow\downarrow\rangle - |\up\downarrow\uparrow\rangle
  \big) \times
    \mathcal S\{|\vec i,\vec j,\vec k\rangle\}
  \bigg\},
  \\
  \nonumber
|e^-,\uparrow\rangle_n =&\ \mathcal S\bigg\{\big(
    2|ddu\rangle-|dud\rangle-|udd\rangle
  \big)
  \\
  & \times \big(
    2|\down\uparrow\uparrow\rangle
    - |\up\uparrow\downarrow\rangle - |\up\downarrow\uparrow\rangle
  \big) \times
    \mathcal A\{|\vec i,\vec j,\vec k\rangle\}
  \bigg\},
  \label{eqn:long_lived_excited_state}
\end{align}
where $\mathcal A\{\cdot\}$ denotes the anti-symmetrization over all six
permutations. The python program identifying the eigenvalues and states
is freely available and referenced in the Data Availability section.
The transition matrix element of the neutron in its ground state is
$5g_\beta/9$ in this model.
The decay rates of the four excited states follow from Fermi's golden rule
as fractions of the ground-state decay rate $\Gamma_g$:
\begin{align}
\Gamma_e &\in \Gamma_g \left\{1,\frac1{25},\frac1{25},\frac9{25}\right\}
\label{eqn:decay_width_ratios}
\end{align}
Only one of the excited neutron states has the same
lifetime under $\beta$ decay as the ground state.
The other three states have lifetimes under $\beta$ decay
that are 25, 25, and $\approx2.78$ times longer, respectively.
This stems from the sensitivity of the spatial wave function of excited states
$|\vec i,\vec j,\vec k\rangle$ to permutations. For instance, $\langle \vec i,\vec k,\vec j|\vec i,\vec j,k\rangle = 0$ for
$\vec j\neq \vec k$, while the spatial wave function of the ground state $|\vec 0,\vec 0,\vec 0\rangle$ is
insensitive to permutations.
Notice that in general the possible initial and final states are not
right and left eigenstates of the interaction Hamiltonian, respectively.
Each of the mutually orthogonal neutron states may decay into
several of the mutually orthogonal proton states.
However, summed over all channels to allowed proton states,
the total $\beta$ decay rate of three of the excited neutron states
is less than but not equal to the decay rate of the ground state.
The case where only two orbitals are distinct $|\vec i,\vec i,\vec j\rangle$ with $\vec i\neq \vec j$ is
similar. The span is two dimensional with
the eigenvalues $\{5,1\}g_\beta/9$ and in general
one excited state has a lifetime less than but not equal to that of
the ground state.

Finally, it is conceivable that isospin symmetry may be broken
in the constituent quark model.
To reflect this, we can replace the flavor wave function part
$2|ddu\rangle-|dud\rangle-|udd\rangle$ by just one of the permutations,
say $|ddu\rangle$.
This has no effect on the ground state; however, the excited states
would generally no longer be eigenstates of the total isospin operator
$\hat I^2$.
For this case, the wave functions constructed from the six permutations of the
spatial part $|\vec i,\vec j,\vec k\rangle$ are all linearly independent and
the eigenvalues of the weak interaction Hamiltonian within this span
are $\{5,2,2,-3,-3,-3\}g_\beta/9$.
Again, only one eigenvalue agrees with the transition element
of the ground state, the other 5 states all have longer lifetimes
under $\beta$ decay.

This model satisfies the condition that the lifetime under $\beta$ decay
of the proposed excited states
is greater than the lifetime of the neutrons in the beam
as stated in \Eqref{eq_ltgamma}
if $\tau_{\gamma}$ is considerably greater than $\tau_\mathrm{beam}$.
Taking, for instance, $|e^-,\uparrow\rangle_n$ of
\Eqref{eqn:long_lived_excited_state} as a possible excited neutron state $\psi_e$,
the difference of the lifetimes $\Delta \tau = \tau_e-\tau_g$ under $\beta$ decay
is $16\tau_g/9$. Assuming no other long-lived excited states are involved,
we can determine the initial conditions using \Eqref{eq_n2ini}.
We find that relatively small abundances of $\psi_e$ states reaching the beam experiment
\be\label{eq_fracn2n1}
\frac{n_e^-(0)}{n_n(0)}=\frac{\Delta \tau}{16\tau_g/9-\Delta\tau}\approx 5.5\times10^{-3},
\ee
are already sufficient to address the N$\lambda$P.
In this model we assume the conservation of spin $\vec S$ in the emission of a $W^-$ boson.
However, the conserved quantity is the total angular momentum, which consists
of spin plus orbital angular momentum $\vec L$
\be
\vec J= \vec S + \vec L.
\ee
Even in the constituent quark model we expect that spin and orbital angular
momentum are not conserved individually, giving rise to transitions
where spatial and spin quantum numbers are both changed by the $\beta$ decay.
This is not considered in the weak interaction Hamiltonian of
\Eqref{eqn:weak_interaction_model}. Spatial transitions in the weak
interaction Hamiltonian, such as
$|\vec i,\vec j,\vec k\rangle$ to $|\vec i',\vec j,\vec k\rangle$, certainly
alter the sensitivity to permutations of the spatial wave functions.
However, we still expect a greater
sensitivity in the case where $(\vec i,\vec j,\vec k)$ are all distinct rather than all equal,
and consequently a smaller overlap between the final and initial excited states.

\subsection{Electromagnetic transition}
\label{sec:electromagnetic_decay}

We finally need to show that the modeled excited states of the neutron may be
sufficiently long lived to reach the beam experiment
but not sufficiently long lived to reach the bottle experiment, i.e.\ that
the inequality
$t_\mathrm{beam} \ll \tau_\gamma \ll t_\mathrm{bottle}$ holds.
We expect a long cascade of electromagnetic transitions between
the neutron in an excited state $\psi_e$ and its ground state $\psi_g$.
Given the wide range of allowed lifetimes and the absence of a reliable model
for the spatial wave functions, we will be satisfied with a rough estimate.
We consider cascades of only electric dipole,
only magnetic dipole, only electric quadrupole, or only magnetic quadrupole
transitions,
but no cascades with mixed transitions for simplicity.

The rate of electric dipole radiation
from an initial state $i$ to a final state $f$ is
\be\label{eq_rate}
P|_{E,d}=\frac{\Delta E_\gamma}{\tau_{f,i}
}=\frac{Z_0^2\alpha \Delta E_\gamma^4}{12 \pi c^2 \hbar^3 }
  |\langle f| \vec r |i \rangle|^2,
\ee
where $\alpha$ is the fine structure constant, $Z_0$ is the fraction of an elementary charge carried by one quark and where
we assume the same charge for all quarks for simplicity.
From here on, we need an estimate of the transition matrix element
$\langle f|\vec r|i\rangle$
and the transition energy $\Delta E_\gamma$ as functions of the quantum numbers.
For the sake of a ballpark estimate, let us assume that the spatial states
are products of solutions of the isotropic three dimensional harmonic oscillator
\be\label{eq_psiHO}
|\vec i,\vec j,\vec k\rangle = \phi_{\vec i}(\vec r_1)\phi_{\vec j}(\vec r_2)\phi_{\vec k}(\vec r_3),
\ee
with
\be
\phi_{\vec i}(\vec r)
= \left( \frac{m \omega}{\pi \hbar}\right)^{3/4}
\prod_{l=1}^3 \frac{1}{\sqrt{ 2^{i_l} i_l!}}
\exp \left[-\frac{m \omega r_l^2}{2 \hbar}
\right] H_{i_l}\left[\sqrt{\frac{m \omega}{\hbar}}r_l\right],
\ee
where $H_i$ are the Hermite polynomials with $i$ nodes.
$\omega$ defines the strength of the effective confining potential
and $m$ the effective mass of a constituent quark,
which we take to be a third of the neutron mass.

The quantum numbers $\vec i,\vec j,\vec k$ determine the number of nodes
in the three cartesian directions of the three quarks, respectively.
For the particular case of the harmonic oscillator, the energy levels are
equidistant and the energy difference between two states
$\Delta E_\gamma= \omega \hbar \Delta i$
depends only on the difference in the total number of nodes $\Delta i$.

The transition matrix elements are non-vanishing only if the total
number of nodes differs exactly by one, say in the first coordinate of the
first quark $i_1$.
For the wave functions of \Eqref{eq_psiHO}, the  squared transition
matrix elements evaluate to
\be\label{eq_transEl}
|\langle (i_1-1,i_2,i_3),\vec j,\vec k| \vec r |\vec i,\vec j,\vec k \rangle|^2\
=\frac{\hbar}{2 m \omega}i_1.
\ee
Inserting \Eqref{eq_transEl} into \Eqref{eq_rate}, one obtains the lifetime
of the $i_1\rightarrow i_1-1$ electric dipole transition
\be
\tau_{i_1}|_{E,d}=\frac{2^3 3 \hbar m \pi }{Z_0^2\alpha i_1 \Delta E_\gamma^2}.
\ee
with $\Delta E_\gamma=\omega\hbar$.
Each electric dipole decay event thus lowers the number of nodes
of a single quark in a single cartesian direction by one
and emits radiation with the same energy $\Delta E_\gamma$.
The events are sequential so the lifetime of the entire decay cascade into
the ground state is the sum of each of the transition lifetimes.
The order of the electric dipole decay events plays no role for the sum.

Assuming that the nuclear production process provides the energy
difference $\Delta E$ between the excited neutron state and its ground state
and that this energy is equally distributed to the three quarks
in three directions, we expect to find 
$I_\mathrm{max}=\Delta E/(9\Delta E_\gamma)$ nodes initially in each coordinate
of each quark.
Thus, the total lifetime of the entire electric dipole decay cascade
from production to the ground state is
\bea \nonumber
\tau_\gamma |_{E,d}&=& 9 \sum_{i=1}^{I_\mathrm{max}} \tau_i|_{E,d}\\
\label{eq_DtEd}
&\approx &\frac{
  2^3 3^2 \pi \hbar m_n \log\left[1+\frac{\Delta E}{9 \Delta E_\gamma}\right]
}{
  \alpha Z_0^2 \Delta E_\gamma^2
},
\eea
where the integer coefficient is rounded to decimal powers.
We approximate electrical quadrupole
transitions by two simultaneous single dipole transitions in different
directions of the same quark. The total lifetime
via electric quadrupole radiation only is found to be
\be
\label{eq_DtEq}
\tau_\gamma |_{E,q}\approx \frac{
  2^6 3^2 5 \hbar m_n^2 \pi \Delta E
}{
  \alpha \Delta E_\gamma^3 (9\Delta E_\gamma+ \Delta E)
}.
\ee

The magnetic moments of the quarks are of the order of two times
the nuclear magnetic moment $\mu_N\approx e \hbar/(2 m_n)$.
For the magnetic quadrupole we used this magnetic moment,
multiplied by the length scale retrieved from the dipole moment of \Eqref{eq_transEl}.
The total lifetimes under only magnetic dipole and only magnetic quadrupole
transitions are respectively
\bea\label{eq_DtMd}
\tau_\gamma |_{M,d}&\approx &\frac{3 c^4 \hbar m_n^2 \Delta E}{\alpha \Delta E_\gamma^4},\\
\label{eq_DtMq}
\tau_\gamma |_{M,q}&\approx &\frac{
  3^7 5 c^6 \hbar m_n^3 \log\left[
    1+\frac{\Delta E}{18 \Delta E_\gamma}
  \right]
}{
  2\alpha \Delta E_\gamma^4
}.
\eea

The multipole radiation rates follow the typical hierarchy:
electric dipole is faster than magnetic dipole radiation and
electric quadrupole is faster than magnetic quadrupole radiation.
Although normally suppressed by the fastest channel, we chose to include
higher moments and magnetic radiation modes into the calculation, since electric dipole
moments are strongly constrained for the ground state of the neutron~\cite{Abel:2020pzs}.
Inverting the relations in Eqs.~(\ref{eq_DtEd}--\ref{eq_DtMq})
allows us to plot the energy estimates $\Delta E_\gamma$ of the emitted photons
as a function of the total lifetime $\tau_\gamma$ of the entire decay cascade
into the ground state.
This plot is shown in Fig.~\ref{fig_Envst}
using a total nuclear excitation energy of the order of $\Delta E\approx 1$\,MeV.

\section{Discussion}
\label{sec_Discussion}

The above results must be discussed from various perspectives. Initially, let us address the most pressing question before delving into more model-specific aspects. The primary question is: ``If excited states exist, why have they gone unnoticed in previous experiments?'' Numerous experiments have explored various aspects of the free neutron and its decay. The most compelling evidence for our hypothesis would undoubtedly be the electromagnetic transition between an excited neutron and a neutron in the ground state. Therefore, our initial focus is on experiments that involve the measurement of direct photons and $\gamma$-rays.

\subsection{Direct $\gamma$ measurements}

The most distinctive property of our hypothesis 
are the transitions from excited states towards the ground state, involving $\gamma$ emission. There has been significant interest in exploring the $\gamma$ background of neutron experiments. A particularly promising series of experiments has been dedicated to constraining the branching ratio of weak decay processes that involve the emission of a $\gamma$ quantum.
\be
n\rightarrow p^+ + e^- + \bar \nu_e + \gamma .
\ee
Such experiments have been performed with beams~\cite{Beck:2002zz,Nico:2006pe,Cooper:2010zza,Cooper:2012me,RDKII:2016lpd}, testing for $\gamma$'s in the broad energy range between $0.4$\,keV and $800$\,keV. 
No evidence for a novel peak in the spectrum was found in this energy range, pushing the branching ratio below $10^{-4}$. This seems to imply a strong constraint on our hypothesis. However, one has to keep in mind the emission of $\gamma$'s from excited neutrons is not correlated to the emission of the charged trigger particle ($e^{-}$ or $p^+$).
Thus, the corresponding events would not be recorded and these bounds of~\cite{Beck:2002zz,Nico:2006pe,Cooper:2010zza,Cooper:2012me,RDKII:2016lpd} can not be imposed directly.
The necessity and method of subtracting the otherwise overwhelming background of uncorrelated $\gamma$'s is nicely explained in~\cite{Cooper:2010zza,RDKII:2016lpd}. 
There is also an experiment that used ultra cold stored neutrons searching for
$\gamma$ events, while not relying on a coincidence trigger~\cite{Tang:2018eln}.
This experiment explored the energy range between $600$\,keV and $1800$\,keV.

\subsection{Mass-sensitive measurements}

Apart from the
$\gamma$ searches,
there is a plethora of experiments 
performed with free neutrons~\cite{Dubbers:2021wqv}.
Each of these experiments has different observables $O$, which require individual detailed quantitative and theoretical analysis. 
However, since at this stage we are interested in orders of magnitude, a semi\-/quantitative estimate is sufficient for now:
Any observable $O(m_n)$ of free neutrons 
is a function of the neutron mass.

An important fact that should be mentioned in the context of the neutron mass is that the neutron mass is measured to extreme precision ($\pm 2\,\mathrm{eV}$)~\cite{ParticleDataGroup:2020ssz}.
However, this measurement involves
deuterium as baryonic bound state~\cite{Kessler:1999zz}. For this reason we formulated the hypothesis f),
that
in the strongly interacting environment
of the baryon, the relatively weak excitation 
of the free neutron can not be formed.
Thus, the excited states of the neutron should also not show up in mass measurements that involve baryonic bound states.

\begin{figure}[hbt]
   \centering
\includegraphics[width=15cm]{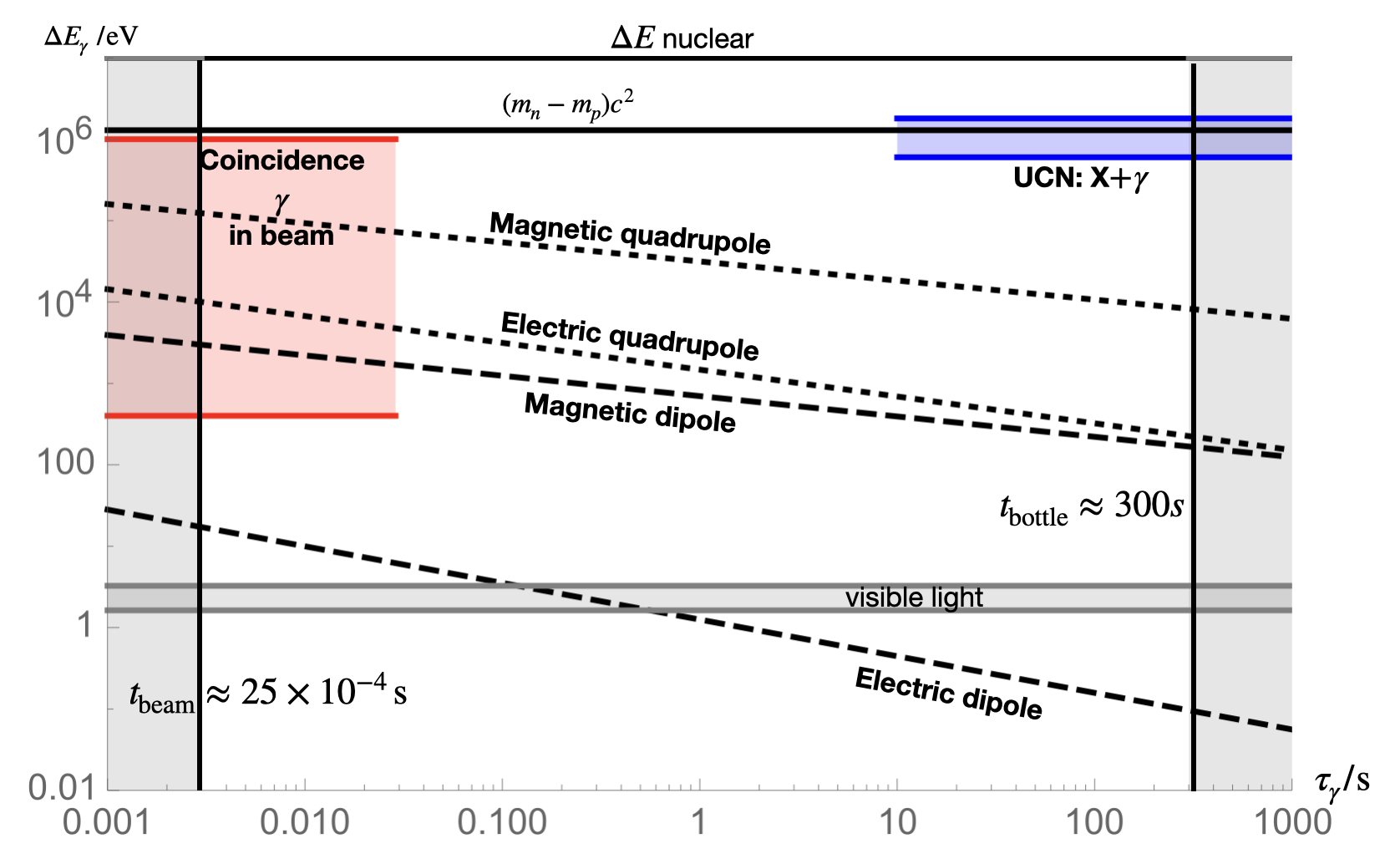}
\caption{\label{fig_Envst}
Electromagnetic radiation energy $\Delta E_\gamma$ versus
total electromagnetic decay times $\tau_\gamma$
of a proposed excited neutron state into the ground state.
For a harmonic-oscillator-like effective confining potential
the frequency of the emitted photons during the decay cascade is a function
of the total decay time.
The dashed and dotted lines show this relation for dipole and quadrupole
radiation, respectively.
The black horizontal lines from top to bottom correspond to:
typical energies available in nuclear fission processes,
the mass difference between neutron and proton, and
the spectrum of visible light.
The red region is excluded by the overlap of $\gamma$ searches for in beam experiments~\cite{Beck:2002zz,Nico:2006pe,Cooper:2010zza,Cooper:2012me,RDKII:2016lpd}, while the blue region is excluded by the $\gamma$ search in the bottle experiment~\cite{Tang:2018eln}. 
White regions are not excluded by the listed observations.
The vertical gray regions on the left and the right are excluded
in the sense that they contradict our hypothesis.
The visible light region is not strictly excluded but questionable,
since it is implausible that excited neutrons emitting visible light would have
gone unnoticed.}
\end{figure}

In the experiments that involve free neutrons, the $m_n$\-/dependence can be typically factored out with some power $P$
\be\label{eq_Pfactor}
O(m_n)=O_0\,  m_n^P.
\ee
Here, $O_0$ is the leftover of the observable, after factorizing the $m_n^P$ dependence. 
If an excited neutron state with a slightly higher mass $m_n'=m_n+ \Delta E/c^2$ existed, it would induce a  shift in the observable $\Delta O$.
The induced relative shift 
in the observable is then
\be
\frac{\Delta O}{O}=P\, \frac{\Delta E}{m_n c^2}.
\ee
If this relative shift would be larger than the relative error ${\mathcal{E}}$ of the experiment, the excited states would have been observed. Thus, imposing ${\mathcal{E}}>\frac{\Delta O}{O}$ one can deduce an upper bound on the excitation energy
\be\label{eq_DeltaEofm}
\Delta E< m_n c^2 \frac{{\mathcal{E}}}{P},
\ee
inflicted by such an observable.
Since the power $P$ only involves changes of order one, the potential of an experiment to constraint $\Delta E$ is dominated by the relative experimental precision ${\mathcal{E}}$. This observation is a belated justification for the simplified factorization~\Eqref{eq_Pfactor}.
The condition \Eqref{eq_DeltaEofm} can be applied to different types of experiments:
\begin{itemize}
    \item Weakly bound wave functions\\
    If the motion of a free neutron is constrained by macroscopic boundary conditions and external potentials, its low energy quantum\-/mechanical wave function is governed by the Schr\"odinger or the Pauli equation. The kinetic term of these equations contains the mass $m_n$ and thus, any observable such as quantum mechanical energy levels or transition rates between states will depend on this mass.
    For example, in the Q\-/bounce experiment the mass dependence of the measured energy levels comes with the power $P=1/3$ and the typical relative precision that can be reached is of the order of ${\mathcal{E}}=10^{-5}$~\cite{Nesvizhevsky:2003ww,Bosina:2023hbf}. 
    Thus, according to \Eqref{eq_DeltaEofm}, these experiments  constrain the 
    $\Delta E$ energies to
    \be \label{eq_DEWB}
    \Delta E_{WB} <3 \times 10^{4}\,\mathrm{eV}.
    \ee
    \item Interference\\
    The evolution of the free neutron wave function can produce interference effects, analogous to the double\-/slit experiment~\cite{Bartosik:2009zz,Cabello:2008zza,Badurek:1976zz}.
    In such epxeriments, the phase of the wave function is proportional to the neutron mass and thus, $P=1$. A typical relative precision that can be reached is of the order of $3\times 10^{-3}$~\cite{Bartosik:2009zz}.
    Thus, according to \Eqref{eq_DeltaEofm}, such interference experiments constrain the energies $\Delta E$ to
    \be \label{eq_DEIE}
    \Delta E_{IE} <3\times 10^{6}\,\mathrm{eV}.
    \ee
    \item Kinematic structure\\
    Exploring the kinematic structure, the neutron decay can further reveal fundamental properties of the standard model of particle physics~\cite{Nesvizhevsky:2003ww}. In the corresponding differential cross sections the dominant $m_n$ dependence is quadratic, which implies $P=2$. The relative theoretical uncertainty is of the order of $10^{-4}$~\cite{Nesvizhevsky:2003ww}.
    Thus, current precision experiments on the kinematic phase space of the neutron decay constrain the energies $\Delta E$ to
    \be \label{eq_DEKS}
    \Delta E_{KS} <5\times 10^{4}\,\mathrm{eV}.
    \ee
\end{itemize}
With these estimates for $\Delta E$ at hand, we still have to remember that
there are actually two possible causes that could explain why these experiments did not see excited neutron states in their measurements.
First, the excitation energy lies below the mass sensitivity of these experiments
as indicated by \Eqref{eq_DeltaEofm}.
Second, the experiments are performed at times $t$ larger than $\tau_\gamma$. 
The quantities $\Delta E_\gamma$ and $\tau_\gamma$ form
a two dimensional phenomenological parameter space.
To exemplify the 
regions not disallowed by existing experiments, we plot the electromagnetic energy versus the timescale~$\tau_\gamma$
in the schematic Fig.~\ref{fig_Envst}.
Even though, this figure is not to be understood as a strict exclusion plot, it provides us a very good intuition of the allowed and tested time and energy scales.
We labeled
this region as ``allowed''.
For comparison, we also plotted the energies that can be expected from our toy, model. Interestingly there is a large overlap between the allowed region and the region preferred by our toy model.

\subsection{Analogies}

The characteristics of an excited state that has a longer lifetime than the ground state, as proposed here for the free neutron, is quite unusual. Nevertheless there are some cases with this characteristics known in nuclear physics. 
Interestingly, nuclear isomers become metastable, if the spin structure between excited and ground state is largely different.
For example, certain nuclear isomers, such as \ce{^{180m}Ta} are stable
in contrast to their ground state~\cite{Lehnert:2016iku}.
Also only recently, other isomers with only tiny excitation energy have been discovered~\cite{Sikorsky:2020peq}.

Another analogy in nuclear physics can be found in the magnetic dipole interactions of currents within atomic nuclei. In these nuclei magnetic dipoles and exchange currents are responsible for the existence of excited states
of otherwise degenerate states with the same nuclear mass~\cite{Heyde:2010ng}. In this context we also want to mention the resonance that was found in polarized neutron--proton scattering~\cite{WASA-at-COSY:2014dmv}.

\subsection{How to test the hypothesis?}

The hypothesis of excited states in the context of the N$\lambda$P can be tested directly.
Such tests would involve designing experiments or observations aimed at detecting the predicted characteristics of these states. Here are several potential approaches to test the hypothesis directly:

\begin{itemize}
    \item[(i)] Perform a beam experiment at later time, which means constructing longer beam pipes with length $L$. 
    Check whether the deduced decay time depends on $L$. This narrows the window between $t_\mathrm{beam}= L/v_n$ and $t_\mathrm{bottle}$ from below.
    \item[(ii)]
    Repeat a bottle experiment at earlier times $t_\mathrm{bottle}$. 
    Check whether the deduced decay time depends on $t_\mathrm{bottle}$. This narrows the window from above. For our toy model, the region between the dashed lines in Fig.~\ref{fig_Envst} favours a relatively large values for $\tau_\gamma$, which increases
    the chances of such an approach.
    \item[(iii)] Search for electromagnetic signatures of the transition between $\psi_e$ and $\psi_g$ along beam pipes, UCN cooling facilities, or before filling UCN bottle containers~\cite{RDKII:2016lpd,Cooper:2010zza,Tang:2018eln}. For example, correlations between $\gamma$ backgrounds from neutron beams have been measured and compared to simulations to a few percent level~\cite{CSNS:2020}.
   Such backgrounds will be of crucial importance. 
   If for instance $\tau_\gamma \approx t_\mathrm{beam}$ then the signature will have to compete with the background of reactor photons. Instead, if  $\tau_\gamma \approx t_\mathrm{bottle}$, the signal will have to compete with the background generated by secondary reactions of the actual weak decays.
 \item[(iv)] Recalculate the angular distributions of the neutron decay products and compare to the nucleon decay parameters measured e.g. in~\cite{Reich:2000au,Klopf:2019afh}.
 \item[(v)] Try to re\-/populate the excited states with fine\-/tuned external radiation.
 \item[(vi)] Use time\-/modulated external radiation to 
 deflect neutrons from a beam and search for a time modulation in the posterior beam flux.
\end{itemize}
Apart from these tests, one can also look for indirect signatures in all sorts of precision experiments with free neutrons, as long as these experiments are performed at $t<t_\mathrm{bottle}$.
Other neutron rich environments 
such as the early universe during Big Bang Nucleosynthesis, or neutron stars, are probably insensitive to the excited states due to the immense red shift of the former and the huge gravitational binding energy of the latter.

One interesting test opportunity we want to mention, is the systematic $3.9\sigma$ energy shift observed in the Q\-/Bounce experiment~\cite{Bosina:2023hbf}. 
Since the energies in Q\-/Bounce depend on the neutron mass, this relative shift of $\approx 10^{-3}$ could be associated to a mass shift due to excited states.

\section{Conclusion}
\label{sec_Conclusion}

In this paper we revisited the neutron lifetime puzzle, assuming that both mutual contradicting experimental results are correct.
Our exploration has led us to propose a novel perspective
on the N$\lambda$P.
The discrepancy could be explained 
by the existence of excited states meeting the conditions (a--f) from Section \ref{sec_Hypo}.
We then presented a consistent
three\-/quark toy model respecting isospin and spin symmetry.
This model satisfies the most important conditions a)--d) quite naturally. 

It is crucial to note that our presented toy model is intended to serve as a starting point for further exploration rather than a definitive explanation. Our primary goal is to emphasize the possibility that the N$\lambda$P may suggest the presence of  excited states with the outlined characteristics. While a substantial portion of our effort went into
constructing and discussing a specific model, we encourage further investigations and alternative approaches to validate and refine our proposed hypothesis of excited states.

In essence, this study opens a pathway for future research to delve deeper into the nature of neutron decay and the potential existence of excited states, providing valuable insights that could contribute to resolving the Neutron Lifetime Puzzle.

\section*{Data Availability}
The python program to identify the eigenvalues and the
neutron and proton eigenstates of the employed
weak interaction model Hamiltonian is freely available and can
be downloaded from \url{https://gitlab.cc4s.org/bumblebee/weak-decay-model-hamiltonian.git}.

\section*{Acknowledgements}

B.K.\ was funded in part by the Austrian Science Fund (FWF)
[P 31702-N27] and [P 33279-N].
We thank H.~Abele and M.~Loewe for detailed comments and we want to thank A.~Santoni, A.~Serebrov, and F.~Wietfeldt for discussions.
We also gratefully acknowledge H.~Skarke for pointing out a mistake with the wave function symmetrization in the previous version of the manuscript.


\end{document}